\documentstyle[]{article}
\font\cero=cmss10 scaled 1728 \font\uno=cmssbx10 scaled 1200
\begin{document}
\begin{flushleft}
{\cero Conformal symmetry of the phase space formulation for
topological string actions} \\[3em]
\end{flushleft}
{\sf R. Cartas-Fuentevilla}\\
{\it Instituto de F\'{\i}sica, Universidad Aut\'onoma de Puebla,
Apartado postal J-48 72570, Puebla Pue., M\'exico
(rcartas@sirio.ifuap.buap.mx).}  \\[3em]

It is proved the conformal invariance of the phase space
formulation for topological string actions associated with the
number of handles and the number of self-intersections of the
world surface. Differences and similarities with the phase space
formulation of an Abelian gauge theory are discussed.\\

\noindent {\uno I. Introduction}
\vspace{1em}

As it is well known, the conformal symmetry plays a crucial role
in the quantum consistency of string theory, and constitutes a way
in which the celebrated critical dimension anomaly occurs.
Specifically, the anomalous scaling behavior at a quantum level
for bosonic string theory disappears only if the background
dimension is 26, and the classical conformal symmetry is thus
preserved. These results have been established considering
basically a string action that is proportional to the area swept
out by the worldsheet.

On the other hand, the topological invariants in Lagrangians for
string theory have been completely neglected because they do not
give dynamics to the string, and additionally  such topological
actions only contribute as global factors in the path integral
formulation of the theory. Thus, one may think that the
topological terms may have not relevant quantum effects, and
particularly will have not effects on the quantum consistency of
the theory.

However, this is only illusory, since recently \cite{1,2} it has
been shown that the topological terms have a dramatic effect on
the phase space formulation for the theory, and even have, by
themselves, a nontrivial phase space formulation, which mimics in
fact the symplectic structure of an Abelian gauge theory \cite{3}.
Consequently the topological invariants will have by themselves a
nontrivial quantum field theory, despite having trivial classical
dynamics.

With these preliminaries, it is natural to ask on the role that
the conformal invariance will play in the quantum aspects of
topological string actions. In this work we want to show, as a
first step in that direction, that the phase space formulation for
these topological actions preserves the conformal symmetry.
Considering the analogy established with an Abelian gauge theory,
we make additionally a comparison of the roles that the conformal
invariance plays in both cases. For this purpose, this work is
organized as follows.

In the next section we set up the basic ideas on a conformal
transformation of the background metric and related formulas in
the imbedding supported background tensor approach for the
differential geometry of an imbedded surface developed by Carter
in \cite{4}. In Section III we out line the conformal properties
of an Abelian gauge theory, with the purpose of making a
comparison with the case of a topological string action developed
in Section IV. We finish in Section V with some remarks and
prospects for the future. \\[2em]

\noindent {\uno II. Conformal transformation}
\vspace{1em}

The general conformal transformation of the background metric
\begin{equation}
     g_{\mu\nu} \rightarrow e^{-2\sigma} \ g_{\mu\nu},
\end{equation}
where $\sigma$ is a local arbitrary function, induces a shift on
the corresponding covariant derivative acting on a vector field
$V_{\nu}$,
\begin{equation}
     \nabla_{\mu} \ V_{\nu} \rightarrow \nabla_{\mu} \ V_{\nu} -
     L^{\lambda}_{\mu\nu} \ V_{\lambda},
\end{equation}
where
\begin{equation}
     L^{\lambda}_{\mu\nu} = -2 \delta^{\lambda}{_{(\mu}} \nabla_{\nu
     )} \ \sigma + g_{\mu\nu} \ \nabla^{\lambda} \ \sigma,
\end{equation}
which is symmetric in $(\mu\nu)$. Additionally (1) induces a
change on the fields with support confined to the world surface
\cite{4}
\begin{eqnarray}
     n^{\mu\nu} \!\! & \rightarrow & \!\! e^{2\sigma}
     n^{\mu\nu}, \quad \quad \bot^{\mu\nu} \rightarrow
     e^{2\sigma} \bot^{\mu\nu}, \nonumber \\
     i_{A}{^{\mu}} \!\! & \rightarrow & \!\! e^{\sigma} \
     i_{A}{^{\mu}}, \quad \quad {\cal E}^{\mu\nu}
     \rightarrow e^{2\sigma} {\cal E}^{\mu\nu},
\end{eqnarray}
where $n^{\mu\nu}$ is the (first) fundamental tensor of the
world-surface, and $\bot^{\mu\nu}$ its complementary orthogonal
projection. Additionally the frame vectors $\{ i^{\mu}_{A} \}$ are
tangential to the world-surface and ${\cal E}^{\mu\nu} = {\cal
E}^{[\mu\nu]} = {\cal E}^{AB} i^{\mu}_{A} \ i^{\nu}_{B}$, where
${\cal E}^{AB}$ is the constant components of the standard
two-dimensional flat space alternating tensor.

Obviously $n^{\mu}_{\nu} \rightarrow n^{\mu}_{\nu},
 \bot^{\mu}_{\nu} \rightarrow \bot^{\mu}_{\nu},$ and ${\cal E
}^{\mu}{_{\nu}} \rightarrow {\cal E}^{\mu}{_{\nu}}$. Considering
the definition of the internal gauge connection \cite{4}
\[
     \rho_{\lambda}{^{\mu}}_{\nu} = n^{\mu\alpha} \ i^{A}_{\nu} \
     \overline{\nabla}_{\lambda} \ i_{\alpha A}, \quad
     \overline{\nabla}_{\lambda} = n^{\alpha}_{\lambda}
     \nabla_{\alpha},
\]
and using Eqs.\ (2)-(4), it is very easy to find the shift of
$\rho_{\lambda}{^{\mu}}_{\nu}$ under the transformation (1):
\begin{equation}
     \rho_{\lambda}{^{\mu}}_{\nu} \rightarrow
     \rho_{\lambda}{^{\mu}}_{\nu} + 2 n_{\lambda [\nu} \
     \overline{\nabla}^{\mu ]} \ \sigma,
\end{equation}
and thus, considering that $\rho_{\lambda} =
\rho_{\lambda}{^{\mu}}_{\nu} {\cal E}^{\nu}{_{\mu}}$, we have
\begin{equation}
     \rho_{\lambda} \rightarrow \rho_{\lambda} - 2
     {\cal E}^{\nu}{_{\lambda}} \ \overline{\nabla}_{\nu} \
     \sigma.
\end{equation}

Similarly, considering the expression $R = \overline{\nabla}_{\mu}
({\cal E}^{\mu\nu} \rho_{\nu}) = {\cal E}^{\mu\nu}
\overline{\nabla}_{\mu} \ \rho_{\nu}$ for the internal scalar
curvature of the two-dimensional world surface \cite{4}, we can
find that
\begin{equation}
     R \rightarrow e^{2\sigma} [R + 2 \overline{\nabla}_{\mu} \
     \overline{\nabla}^{\mu} \ \sigma],
\end{equation}
which implies that the Euler characteristic of the embedding
two-surface \cite{4}
\begin{equation}
     \chi = \int \sqrt{-\gamma} \ R \ d\overline{\Sigma},
\end{equation}
related geometrically with the number of handles of the
world-surface, undergoes the transformation
\begin{equation}
     \chi = \chi + 2 \int \sqrt{-\gamma} \ \overline{\nabla}_{\mu} \
     \overline{\nabla}^{\mu} \ \sigma \ d\overline{\Sigma},
\end{equation}
and then $\chi$ is, modulo a total divergence, conformally
invariant, as it is well known in the literature. In Eq.\ (9) we
have considered that $\sqrt{-\gamma} \rightarrow e^{-2\sigma} \
\sqrt{-\gamma}$. \\[2em]

\noindent {\uno III. Conformal symmetry in an Abelian gauge
theory}
\vspace{1em}

As it is well known in the literature, the Abelian gauge theory
described by the equations
\begin{eqnarray}
     \nabla^{\mu} \ F_{\mu\nu} \!\! & = & \!\! 0, \\
     \nabla_{\{\mu} \ F_{\alpha\beta\}} \!\! & = & \!\! 0,
\end{eqnarray}
is conformally invariant only if the background spacetime
dimension and the {\it conformal weight} of $F_{\mu\nu}$ are fixed
to be 4 and 0 respectively. Additionally, if we set up the
conformal weight of $A_{\nu}$ to be 0, the covariant and
gauge-covariant symplectic structure for the theory given by
\cite{5}
\begin{equation}
     \widehat{\omega} = \int_{\Sigma} \delta F^{\mu\nu} \ \delta
     A_{\nu} \ d \Sigma_{\nu},
\end{equation}
where $\Sigma$ is a spacelike hypersurface, is also conformally
invariant. Under these conditions, the left-hand side of Eq.\ (10)
transforms {\it homogeneously} under (1),
\begin{equation}
     (\nabla^{\mu} F_{\mu\nu})' = e^{2\sigma} \ \nabla^{\mu}
     F_{\mu\nu},
\end{equation}
which implies that if (and only if) the Eq.\ (10) is satisfied,
the conformally transformed version also is.

Similarly, the left-hand side of Eq.\ (11) transforms as
\begin{equation}
     (\nabla_{[\mu} F_{\alpha\beta ]})' = \nabla_{[\mu}
     F_{\alpha\beta ]}.
\end{equation}
\\[1em]

\noindent {\uno IV. Conformal symmetry of the phase space
formulation for $\chi$}
\vspace{1em}

As shown in \cite{3}, the phase space formulation for the
topological string action $\chi$ is given in terms of the
equations
\begin{eqnarray}
     \overline{\nabla}^{\mu} \ {\cal E}_{\mu\nu} \!\! & = & \!\!
     0, \\
     n^{\sigma}_{\{\beta} n^{\rho}_{\gamma} \overline{\nabla}_{\alpha
     \}} (R \ {\cal E}_{\sigma\rho}) \!\! & = & \!\! 0,
\end{eqnarray}
and the covariant and gauge-invariant symplectic structure given
by
\begin{equation}
     \omega = \int \delta (\sqrt{-\gamma} \
     {\cal E}^{\mu\nu}) \ \delta \ \rho_{\nu} \ d
     \overline{\Sigma}_{\mu},
\end{equation}
which mimic entirely the mathematical structure of that of the
Abelian gauge theory given in terms of Eqs.\ (10), (11), (12), in
the indicated order \cite{3}.

As we shall see in this section, despite the close analogy between
the phase space formulation of both theories, there exist
important differences in the conditions required for ensuring the
conformal invariance of those formulations.

We suppose first that we do not know the conformal weight of
${\cal E}_{\mu\nu}$ (-2, according to (Eq.\ (4)), and to consider
that in general
\[
     {\cal E}_{\mu\nu} \rightarrow e^{s\sigma} \ {\cal
     E}_{\mu\nu},
\]
and let us prove that the conformal invariance of Eq.\ (15)
requires precisely that conformal weight induced on ${\cal
E}_{\mu\nu}$ by the transformation (1).

Considering the left-hand side of Eq.\ (15) conformally
transformed, and the Eqs.\ (2)-(4), we have
\begin{eqnarray}
     (n^{\mu\alpha})' \ \nabla'_{\alpha} \ {\cal E}'_{\mu\nu} \!\!
     & = & \!\! e^{2\sigma} \ n^{\mu\alpha} [\nabla_{\alpha}
     (e^{s\sigma} {\cal E}_{\mu\nu}) - L^{\lambda}_{\alpha\mu}
     (e^{s\sigma} {\cal E}_{\lambda\nu}) - L^{\lambda}_{\alpha\nu}
     (e^{s\sigma} {\cal E}_{\mu\lambda})], \nonumber \\
     (\overline{\nabla}^{\mu} {\cal E}_{\mu\nu})' \!\! & = & \!\!
     e^{(s+2)\sigma} \ \overline{\nabla}^{\mu} {\cal E}_{\mu\nu} -
     (s+2) e^{(s+2)\sigma} {\cal E}_{\mu\nu} \
     \overline{\nabla}^{\mu} \ \sigma,
\end{eqnarray}
where we have considered the symmetry of $n^{\mu\nu}$, the
antisymmetry of ${\cal E}_{\mu\nu}$, and that $n^{\mu\nu}
n_{\mu\nu} = 2$ for a two-dimensional world surface \cite{4}.

Therefore, Eq.\ (18) shows in a manifest way that if $s = -2$, the
conformal weight naturally induced on ${\cal E}_{\mu\nu}$ by (1),
$\overline{\nabla}^{\mu} {\cal E}_{\mu\nu}$ is strictly a
conformal invariant without any restriction on the background
dimension. In the case of the Abelian gauge theory, $\nabla^{\mu}
F_{\mu\nu}$ is not strictly a conformal invariant (see Eq.\ (13)),
but it transforms {\it homogeneously} under (1). In this manner,
if (and only if) Eq.\ (15) holds, its conformally transformed
version is also satisfied.

On the other hand, Eq.\ (16) corresponds to the {\it closure} of
the two-form $R {\cal E}_{\mu\nu}$, which ensures that locally it
can be written as the {\it exterior derivative}
$(\overline{\partial})$ of the one-form $\rho_{\mu}$ \cite{4},
\begin{equation}
     \overline{\partial} (R {\cal E}) = 0 \quad \Leftrightarrow
     \quad R {\cal E} = \overline{\partial} \rho.
\end{equation}

In passing, we point out a mistake in the equivalence relations
(A.10), and (A.14) in Ref.\ \cite{4}, whose left-hand sides must
be not $\partial\partial F$, and $\overline{\partial}
\overline{\partial} F$, but only $\partial F$ (or
$\partial\partial A$), and $\overline{\partial} F$ (or
$\overline{\partial}\overline{\partial} A$) respectively. In
components, the right-hand side of the equivalence relation (19)
is expressed as \cite{4}
\begin{equation}
     R {\cal E}_{\mu\nu} = (\overline{\partial} \rho)_{\mu\nu} = 2
     n^{\sigma}_{[\nu} \overline{\nabla}_{\mu ]} \ \rho_{\sigma},
\end{equation}
and Eq.\ (16) corresponds, in components, to the left-hand side of
(19).

Let us prove that this closure-exactness property is preserved
under (1). Considering Eqs.\ (2), (6), and (20), $R {\cal E}$
undergoes the transformation
\begin{equation}
     (R {\cal E}_{\mu\nu})' = R {\cal E}_{\mu\nu} + 2 n^{\alpha}_{[\mu}
     \overline{\nabla}_{\mu ]} \ B_{\alpha},
\end{equation}
where $B_{\alpha} = 2 \overline{\nabla}_{\nu} ({\cal E}_{\alpha}
{^{\nu}} \sigma)$ is the conformal shift of $\rho_{\mu}$ (see Eq.\
(6)). Equation (21) implies that $R {\cal E}$ changes by the
exterior derivative of $B$;
\begin{equation}
     (\overline{\partial} B)_{\mu\nu} = 2 n^{\alpha}_{[\mu}
     \overline{\nabla}_{\mu ]} \ B_{\alpha}.
\end{equation}

Therefore, considering Eqs.\ (2), and (21) we have that the
left-hand side of (19) conformally transformed is given by
\begin{equation}
     [n^{\sigma}_{\{\beta} n^{\rho}_{\gamma} \overline{\nabla}_{\alpha\}}
     (R {\cal E}_{\sigma\rho})]' = n^{\sigma}_{\{\beta}
     n^{\rho}_{\gamma} n^{\mu}_{\alpha\}} \nabla'_{\mu}
     (R {\cal E}_{\sigma\rho})' = n^{\sigma}_{\{\beta}
     n^{\rho}_{\gamma} \overline{\nabla}_{\alpha\}}
     (R {\cal E}_{\sigma\rho}) + n^{\sigma}_{\{\beta}
     n^{\rho}_{\gamma} \overline{\nabla}_{\alpha\}} [2
     n^{\lambda}_{[\sigma} \overline{\nabla}_{\rho ]}
     B_{\lambda}],
\end{equation}
where the terms proportional to $L^{\lambda}_{\mu\nu}$ vanish as a
consequence of the symmetry of $L^{\lambda}_{\mu\nu}$, and the
antisymmetry of ${\cal E}_{\mu\nu}$ and of the exterior derivative
of $B$ (22). Equation (23) corresponds, in components, to the
identity
\begin{equation}
     \overline{\partial}' (R {\cal E})' = \overline{\partial} (R {\cal
     E}) + \overline{\partial} \overline{\partial} B.
\end{equation}

The last term in (24) vanishes identically because of the
nilpotency of the exterior derivative $\overline{\partial}^{2}=0$
\cite{4} , and then $\overline{\partial}(R {\cal E})$ is a strict
conformal invariant (such as the case of an Abelian gauge theory,
Eq.\ (14)), and its vanishing is guaranteed under a conformal
transformation.

We analyze now the conformal symmetry of the symplectic structure
(17). Considering that ${\cal E}^{\mu\nu} \rightarrow e^{2\sigma}
{\cal E}^{\mu\nu}$, and $\sqrt{-\gamma} \rightarrow e^{-2\sigma}
\sqrt{-\gamma}$, then $\sqrt{-\gamma} \ {\cal E}^{\mu\nu}$ turns
out to be a conformal invariant, and consequently its variation
$\delta (\sqrt{-\gamma} {}\cal E^{\mu\nu})$ is also a strict
conformal invariant. On the other hand, the conformal shift on the
connection $\rho_{\lambda}$ in Eq.\ (6), induces a corresponding
shift on its variation given by
\begin{equation}
     \delta \rho_{\lambda} \rightarrow \delta \rho_{\lambda} - 2
     {\cal E}^{\nu}{_{\lambda}} \ \overline{\nabla}_{\nu} \
     \sigma,
\end{equation}
and then the conformal shift on the symplectic structure $\omega$
is given by
\begin{equation}
     \omega' = \omega + \int {\cal E}_{\nu}{^{\alpha}} \ \delta
     (\sqrt{-\gamma} \ {\cal E}^{\mu\nu}) \
     \overline{\nabla}_{\alpha} \ \sigma \ d
     \overline{\Sigma}_{\mu}.
\end{equation}

Let us prove now that the second term on the right-hand side is a
total divergence. Considering the symmetry $ {\cal
E}_{\nu}{^{\alpha}} \ \delta (\sqrt{-\gamma} \ {\cal  E}^{\mu\nu})
=  {\cal  E}_{\nu}{^{\mu}} \ \delta (\sqrt{-\gamma} \ {\cal
E}^{\alpha\nu})$, and that $\overline{\nabla}_{\mu} [\delta
(\sqrt{-\gamma} \ {\cal  E}^{\mu\nu})] = 0$ \cite{3}, such a term
can be rewritten as
\begin{eqnarray}
     \int {\cal  E}_{\nu}{^{\alpha}} \ \delta (\sqrt{-\gamma} \
     {\cal  E}^{\mu\nu}) \overline{\nabla}_{\alpha} \ \sigma \ d
     \overline{\Sigma}_{\mu} \!\! & = & \!\! \int
     {\cal  E}_{\nu}{^{\mu}} \ \overline{\nabla}_{\alpha} [\delta
     (\sqrt{-\gamma} \ {\cal  E}^{\alpha\nu}) \ \sigma ] \ d
     \overline{\Sigma}_{\mu} \\
     \!\! & = & \!\! \int_{\Sigma} \overline{\nabla}_{\alpha}
     [{\cal  E}_{\nu}{^{\mu}} \ \delta (\sqrt{-\gamma} \
     {\cal  E}^{\alpha\nu}) \ \sigma ] \ d \overline{\Sigma}_{\mu} -
     \int_{\Sigma} \sigma \ \delta (\sqrt{-\gamma} \
     {\cal  E}^{\alpha\nu}) (\overline{\nabla}_{\alpha} \
     {\cal  E}_{\nu}{^{\mu}}) d \overline{\Sigma}_{\mu}, \nonumber
\end{eqnarray}
where the integrand of the last term vanishes, because
$\overline{\nabla}_{\alpha} \ {\cal  E}_{\nu}{^{\mu}} =
K_{\alpha\tau}{^{[\mu}} {\cal  E}_{\nu ]}$ and then
$K_{\alpha\tau}{^{\mu}} d \overline{\Sigma}_{\mu} = 0$, and
$\delta (\sqrt{-\gamma} \ {\cal  E}^{\mu\nu}) K_{\alpha\tau\nu} =
0$ ($K_{\alpha\tau}{^{\mu}}$ is orthogonal in its last indice, and
$d\overline{\Sigma}_{\mu}$ and $\delta (\sqrt{-\gamma} \ {\cal
E}^{\mu\nu})$ are tangential in their indices). Therefore, the
symplectic structure changes by a total derivative, such as the
action $\chi$ itself (see Eq.\ (9)), under a conformal
transformation, and imposing the appropriate compactness
properties on the field variations at the boundary $\partial
\Sigma$, we have finally that $\omega$ is a conformal invariant,
without any restriction on the background dimension. \\[2em]

\noindent {\uno V. Remarks and prospects}
\vspace{1em}

In this manner, considering the basic idea of a phase space
formulation of preserving the relevant symmetries of the
theory\cite{5}, the conformal symmetry of the action $\chi$ is
preserved in its corresponding canonical formalism given by Eqs.\
(15), (16), and (17), without any restriction.

As it is well known in (bosonic) string theory, the classical
symmetry is preserved in the quantum domain only if the background
dimension is fixed to be $26$. One can ask if $\chi$ has a
relevant effect on this particular question, since the
quantization of $\chi$ can be achieved, in principle, using the
phase space formulation studied here. Works along these lines are
in progress, and will be subject of future communications.

There exists another topological invariant associated with the
two-dimensional world surface, and related geometrically with the
number of self-intersections of such a surface. The corresponding
phase space formulation proves to have also the form of an Abelian
gauge theory \cite{3}, and preserves the conformal symmetry of the
original action in an entirely similar way to the case developed
here. However, this case is from the beginning dimensionally
restricted, since the topological invariance of the action
requires a $4$-dimensional background. This dimensional
restriction makes its analogy with an Abelian gauge theory closer
than that of $\chi$, in accordance with the results presented in
Sec. III. \\[2em]

\begin{center}
{\uno ACKNOWLEDGMENTS}
\end{center}
\vspace{1em}

The author acknowledges the support from the Sistema Nacional de
Investigadores (M\'exico). \\


\begin{thebibliography}{}

\setlength{\itemsep}{-.50em}
\bibitem{1} R. Cartas-Fuentevilla, J.\ Math.\ Phys., {\bf 45}, 602
(2004), math-ph/0404004.
\bibitem{2} R. Cartas-Fuentevilla, and A. Escalante, {\it Topological terms
and the global symplectic geometry of the phase space in string
theory}, Trends in \ Math.\ Phys., Nova publishing, to be
published (2004),  math-ph/0404001.
\bibitem{3}R. Cartas-Fuentevilla, {\it Fluctuating topological
invariants in string theory as an Abelian gauge theory}, submitted
to Phys.\ Lett.\ B., (2004), math-ph/0404011.
\bibitem{4} B. Carter, J.\ Geom.\ Phys., {\bf 8}, 53 (1992).
\bibitem{5} C. Crncovi\'c and E. Witten, in {\it Three Hundred Years
of Gravitation}, edited by S. W. Hawking and W. Israel (Cambridge
University Press. Cambridge, 1987).
\end{thebibliography}
\end{document}